\documentclass[conference]{IEEEtran}
\IEEEoverridecommandlockouts
\usepackage{cite}
\usepackage{amsmath,amssymb,amsfonts}
\usepackage{algorithmic}
\usepackage{graphicx}
\usepackage{textcomp}
\usepackage{xcolor}
\usepackage{hyperref}
\def\BibTeX{{\rm B\kern-.05em{\sc i\kern-.025em b}\kern-.08em
    T\kern-.1667em\lower.7ex\hbox{E}\kern-.125emX}}
\begin{document}

\title{Optimal Transmission Topology for Facilitating the Growth of Renewable Power Generation\\
\thanks{This project has received funding from the European Union's Horizon 2020 research and innovation programme under grant agreement No 773406, project OSMOSE (Optimal System-Mix Of flexibility Solutions for European electricity).}
}

\author{\IEEEauthorblockN{Emily Little*\textsuperscript{$\ddagger$}, Sandrine Bortolotti*, Jean-Yves Bourmaud*, Efthymios Karangelos\textsuperscript{$\dagger$}, Yannick Perez\textsuperscript{$\ddagger$}}
\IEEEauthorblockA{
*RTE R\&D, Paris La D\'{e}fense, France \\
\textsuperscript{$\dagger$}Montefiore Institute, University of Li\`{e}ge, Li\`{e}ge, Belgium \\
\textsuperscript{$\ddagger$}Laboratoire de G\'{e}nie Industriel, CentraleSup\'{e}lec, Gif-sur-Yvette, France}
}
\IEEEoverridecommandlockouts
\IEEEpubid{\makebox[\columnwidth]{978-1-6654-3597-0/21/\$31.00~\copyright2021 IEEE \hfill} \hspace{\columnsep}\makebox[\columnwidth]{ }}

\maketitle

\IEEEpubidadjcol

\begin{abstract}
Transmission topology control is a tool used by system operators in the role of a control action taken into account as a preventive or corrective action relative to a specific outage or set of outages. However, their inclusion in most electricity market frameworks is limited. With the increasing penetration of intermittent energy sources, optimal topology can be used as a lever of flexibility to decrease the total system cost. This paper demonstrates the evolution of optimal topology control on systems with increasing quantities of intermittent renewable energy along two axes. First, the effects of the increased variable sources on the variations of optimal topology are explored. Second, we elaborate on the growing advantages of exploiting transmission-level grid flexibility in terms of total system cost. Case studies are performed on a modified RTS-96 network.
\end{abstract}


\section{Introduction} \label{Intro}

With increasing penetration of renewable energies, more and more focus is being placed on flexibility in the electricity system. Improved flexibility is required not only on the generation and consumption sides of the system, but also in the transport, which is crucial to achieve the secure physical execution of the electricity market. Along with costly remedial actions such as the redispatch of generation units, these possibilities include topological changes (which itself includes transmission switching and bus splitting and merging, among others), phase shifting transformers (PSTs) and high voltage direct current (HVDC) lines. 

While power flow control devices and redispatch and even some topology control\footnote{The extent to which system operators use topology control varies widely case by case and often lacks transparency. Reference \cite{hedman_review_2011} elaborates on the operational parameters for some system operators.} are regularly used by system operators (Transmission System Operators (TSOs), Independent System Operators (ISOs), etc.), their integration in the electricity market framework remains incomplete.  Their usage remains largely reserved for ensuring the security of the grid close to real time and is generally determined individually by system operators. However, as the inherent system variability increases, these tools will become more and more beneficial, not only from a security perspective, but also economically. Their integration into current market paradigms presents certain challenges that will likely require fundamental market design changes. This article aims to quantify the increasing advantage that arises through market integration of these levers by way of analysis of a basic case study.

From a mathematical perspective, control of HVDC lines, phase shifting transformers (PSTs) as well as other power flow devices that operate through line reactance control (FACTs) can require binary variables for their full modelization, as in \cite{djelassi_hierarchical_2018}. However, they are often approximated through the use of continuous variables, facilitating their integration into a variety of optimization problems \cite{van_den_bergh_dc_2014}. Optimal topology control can not be approximated in this way. 

The optimal transmission switching (OTS) problem has been well-covered in the literature, beginning in the early 1980s. As described by Mazi et al. \cite{mazi_corrective_1986} at the time: "An experiment was conducted in which a power system model was set up to reflect a contingency condition resulting in a branch overload. A corrective action for this overload was calculated using a linear programming solution and resulted in a shift in generation. The same contingency condition was then presented to a system operator on a dispatcher training simulator… The operator's initial response to the overload was to simply switch the network. This relieved the overload and allowed the generation to remain at economic loading." Mazi et al. then point out that while transmission switching is an action often taken by network operators, it is not often automated or pre-calculated. 

Certain aspects of this remain true today. Topology control poses a difficult problem, both in its resolution and the question of transparency to market players. The research on this topic can be split roughly into two large blocks. The first period from the early 1980s to the early 2000s (largely before the liberalisation of electricity markets) focused almost exclusively on the technical aspects of the problem. Reference \cite{van_amerongen_security_1980} introduced the problem as a method to correct overloads after N-1 events. It was found early on that the problem was quite complex\footnote{See \cite{lehmann_complexity_2014} for in depth complexity analysis of the problem.}  and thus the large part of the research focused on fast heuristics, decomposition and other methods to reduce the search space. 

Reference \cite{rolim_study_1999} performed a literature review in 1999 and found optimization problem search spaces limited by: operator experience \cite{bakirtzis_incorporation_1987} \cite{schnyder_security_1990}, sensitivity factors \cite{bacher_network_1986} \cite{mazi_corrective_1986}, off-line analysis \cite{koglin_corrective_1985} \cite{bertram_integrated_1990}, branch and bound heuristics \cite{gorenstin_b_framework_1987}, loop flow analysis \cite{freitas_e_silva_switching_1993}, Z-matrices \cite{makram_selection_1989}, electrical distance, and development of specific strategies based on flow direction. Up to this point, reference \cite{rolim_study_1999} found that the most common objectives of optimal transmission switching were solving branch overloads and voltage problems \cite{bakirtzis_incorporation_1987} \cite{rolim_secte_1995}. Some studies also looked at minimizing losses \cite{dodu_search_1981} \cite{schnyder_security_1990} and some preventive security enhancement \cite{glavitsch_h_combined_1984} \cite{schnyder_security_1990}. They found that the large majority of the literature concentrated on corrective remedial actions, rather than preventive. 

While this type of research continues to this day, a new block of research began developing in the early 2000s analyzing the economic aspects of the problem. This was driven by several factors. Firstly, the somewhat global liberalization of electricity markets brought about a reframing of the context of the problem itself. In addition, in the past, the added value from automation of grid flexibility was limited as the flows resulting from conventional power production sources were relatively stable and well-known. The addition of variable renewable energies brought a renewed interest in this area as power flows become more unpredictable and irregular (not to mention more highly-distributed farther from load centers, where power lines were previously dimensioned for smaller and more regular flows). 

Reference \cite{oneill_dispatchable_2005} proposed dispatchable transmission asset auctions. A series of papers co-optimizing unit commitment or economic dispatch and optimal transmission switching, published between 2009 and 2012, found gains of between 3-25\% of total cost due to transmission switching \cite{fisher_optimal_2008} \cite{hedman_co-optimization_2009} \cite{hedman_review_2011}. Reference\cite{hedman_co-optimization_2009} used the $B\theta$ form of the DC-OPF and limited the total number of switches allowed by the algorithm to make the problem tractable. This method was expanded to include N-1 contingencies in \cite{hedman_optimal_2009}. Reference \cite{goldis_shift_2017} developed a PTDF approach to improve the computational time of the co-optimized problem. Often in these studies, the problem was decomposed, iterating between a unit commitment with fixed topology and an optimal transmission switching problem with fixed generation \cite{hedman_review_2011}. Reference \cite{han_impacts_2016} found 0.3–2.03\% reduction in total cost in zonal electricity markets using a similar decomposition.

Beyond the obvious gains of optimizing the grid topology for different dispatch profiles, \cite{hedman_optimal_2008}, \cite{hedman_optimal_2011} and \cite{oneill_dispatchable_2005} analyzed the specific economic sensitivities related to optimal transmission switching. These authors found that the total gain from optimal transmission switching could be approached with a limited number of switching operations. However, solutions with almost identical total savings led to large differences in both LMPs (Locational Marginal Prices, also referred to as nodal prices) and the wealth transfers between market participants. Since the problem is quite large and unlikely to attain an exact optimum in operation, this poses a problem for transparency to market players. In addition, the changes in topology can lead to revenue inadequacy regarding the long-term FTR (Financial Transmission Rights) market. The implication is that a reconsideration of LMP-based market trading would be useful to exploit the benefits of optimizing the transmission topology. Similar sensitivities of zonal prices were found by \cite{mekonnen_influence_2012} and \cite{mekonnen_power_2013} of the flow-based market coupling to PST and HVDC parameters. 

To summarize, the technical and economic benefits of optimal transmission switching are established in the literature. The goal of this paper is to expand the existing studies by focusing on a current aspect, the need to accommodate increasing amounts of variable renewable power generation through the transmission system. Unlike the optimal transmission switching analyses performed by \cite{hedman_co-optimization_2009} and \cite{hedman_optimal_2008}, this study focuses on bus splitting and merging rather than switching branches in and out of service, which are often more complicated from an operational perspective \cite{goldis_shift_2017}. Specifically, the paper studies how the use and economic value of transmission switching increase with the growth of renewable power generation. 

Section \ref{Methodology} explicits the methodology used for this study, followed by a description of the case study in Section \ref{CaseStudy} and results in Section \ref{Results}. Finally, the next steps in this research are described in Section \ref{FurtherResearch}.

\section{Methodology} \label{Methodology}
The aim of this study is to analyze optimal topology control on systems with increasing penetration of variable renewable energies. More specifically, we aim to assess two facets of this topic: firstly, how increasing system variability calls for the usage of a wider variety of grid flexibility actions; and secondly how the economic benefits of grid flexibility evolve as the generation side of the electricity system becomes more variable.

In order to analyze these two aspects, a full year study was run with a DC-OPF performed on each hour of the year. Several scenarios with increasing wind production were generated and the results from three variations of the DC-OPF were compared: 1) a DC-OPF without flexible topology; 2) a DC-OPF with optimal topology control (OTC); and 3) a generation economic dispatch, ignoring the transmission network. The no-network (or copper plate) variation gives a lower limit of the total production cost and when compared to the first variation, assigns a value to the added system cost due specifically to network constraints. The second variation, with OTC, allows for analysis of how the expansion of use and variability of optimal grid flexibility actions could benefit the customer in future energy systems.

The lossless DC-OPF formulation used for this study includes the nodal balance constraints, generation limits, and branch flow limits \cite{fisher_optimal_2008}. Constraints imposing full grid connectivity were also included as shown in \cite{djelassi_hierarchical_2018}. For the DC-OPF with OTC, the following constraints were added for each breaker with the possibility to be split: 

\indent $\forall k\in K_{B}:$
\begin{gather}
    -(1-\delta_{k})M_B \leq \theta_f - \theta_t \leq (1-\delta_{k})M_B\\
    -\delta_{k}M_B \leq P_{fl_{k}} \leq \delta_{k}M_B
\end{gather}
where $K_{B}$ is the set of controllable breakers, $M_B$ is a parameter sufficiently large to render the inequality ineffective according to the breaker status variable, $\delta_{k}$ is the binary variable representing the breaker open/closed status, $P_{fl_{k}}$ is the flow on branch k, and $\theta_f$ and $\theta_t$ are the voltage angles on the origin and end nodes, respectively. While the majority of control actions considered in the case study were splitting and merging buses, a small set of controllable branches are included as well. To that end, the following constraints were included for each branch that can be open or closed:

\indent $\forall k\in K_{CB}:$
\begin{gather}
    -(1-z_k)M_K \leq -P_{fl_{k}} + B_{k}(\theta_f - \theta_t) \leq (1-z_k)M_K \\
    -P_{fl_{k}}^{max}z_k \leq P_{fl_{k}} \leq P_{fl_{k}}^{max}z_k
\end{gather}
where $K_{CB}$ is the set of controllable branches, $z_k$ is the binary variable representing the branch open/closed status, $M_K$ is a parameter similar to $M_B$ described above and $B_{k}$ is the susceptance. In order to assess the idealized system flexibility, no limits are placed on the number of control actions that can be used.

The objective function minimizes the total production cost. For the DC-OPF with OTC, a small wear-and-tear cost was also added to the objective function for each line that is opened or bus that is split.\footnote{Note that this penalty cost was not included in the total production cost calculation.} The possibility for load shedding at a very high cost was included in the objective function for all three DC-OPF variations. The objective function is as follows\footnote{Note that the two middle terms of Equation \ref{eqn:objfxn} are included only in the DC-OPF with OTC.}: 

\begin{multline} \label{eqn:objfxn}
\min \sum_{g\in G}{c_gP_g} + c_{w}(N_B - \sum_{k\in K_{B}}{\delta_k}) + \\c_{w}(N_{CB} - \sum_{k\in K_{CB}}{z_k}) + c_{LS}P_d^{LS}
\end{multline} 
where $G$ is the set of generators with the cost and production represented by $c_g$ and $P_g$. $N_{B}$ and $N_{CB}$  are the number of controllable breakers and branches, $c_w$ is the wear-and-tear cost associated with the control actions and $c_{LS}$ and $P_d^{LS}$ are the load shedding cost and quantities. 

For each scenario and DC-OPF variation, the total system cost and the dispatch were calculated. Finally, the nodal prices were calculated by rerunning a DC-OPF with all binary variables fixed as described in \cite{oneill_efficient_2005}. For the purposes of this study, renewable power generation is considered as a free resource (no cost), hence prioritized by the chosen objective function (5). Curtailment of renewables is considered at no additional cost. 

\section{Case Study} \label{CaseStudy}

\subsection{Network and Data Description} \label{Network}
The network used for this study is based on the modified RTS-96 in \cite{barrows_ieee_2020}, which is a modernized version of the original RTS-96 network \cite{grigg_ieee_1999}.\footnote{The data can be found at \href{https://github.com/GridMod/RTS-GMLC}{https://github.com/GridMod/RTS-GMLC}. } Some simplifications were made to this data: namely, the piecewise linear cost model used in the original data set was reduced to a simple marginal price, minimum power output was zero for all plants, and storage was not included as intertemporal constraints were not modeled. 

\begin{figure}
    \centering
    \includegraphics[width=8.8cm]{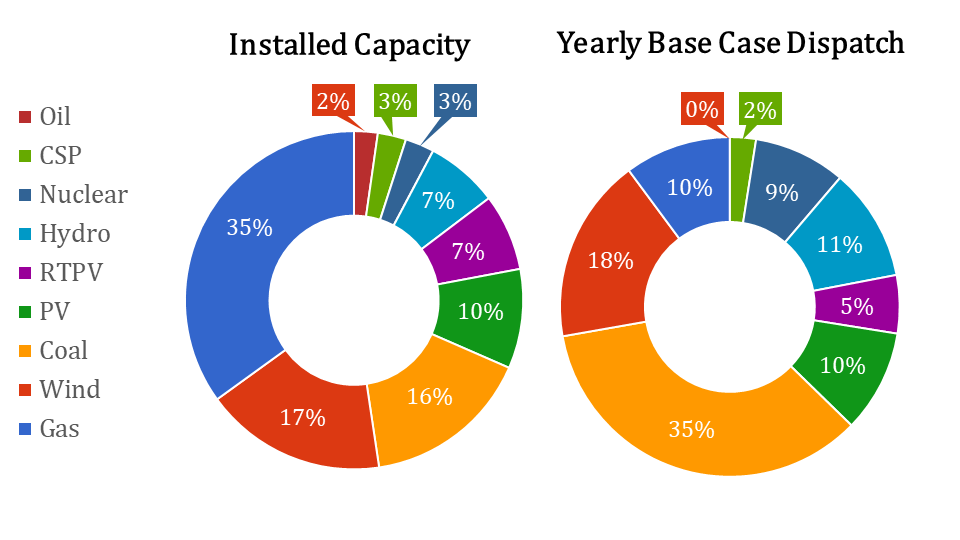}
    \caption{System Generation by Fuel}
    \label{fig:systemGen}
\end{figure}

Reference \cite{barrows_ieee_2020} artificially places the RTS-96 network in the Southwest United States in order to have yearly renewable data. The network is characterized by high load in the summer, fairly steady PV production throughout the year and increased average wind production in the winter months. The installed capacity and energy dispatched in this base case scenario are shown in Figure \ref{fig:systemGen}. 

\subsection{Determining Topology Control Actions}
The modified RTS-96 system did not originally include grid flexibility controls. A subset of the network buses were selected as controllable as well as a few branches. In order to choose which buses would be considered for control actions, an analysis was performed using two seasonally representative months: January and July. All buses were allowed to split or merge and those that changed position more than a certain threshold (set at 15\% of the time) were selected for the case study. 

While the study focuses largely on bus splitting or merging, four lines per zone were additionally considered for transmission switching as they are lines that have a parallel counterpart. In total, 30 topological actions are included in the system. 18 out of the 73 nodes were deemed controllable.

\subsection{Generation Scenarios}
Several scenarios with increasing wind capacities were generated from the base case data described above. We consider that the resource at each of the plant locations had not yet been fully utilized and could thus be increased up to three times that of the base case. Twelve scenarios were generated wherein all wind power plants increase their power output proportionally from 0-300\% of the base case capacity, or 0-7.52 GW of installed capacity.  

\section{Results} \label{Results}

\subsection{Effect of Increasing Generation Variability on Usage of Grid Flexibility Controls}
\begin{figure}
    \centering
    \includegraphics[width=8.8cm]{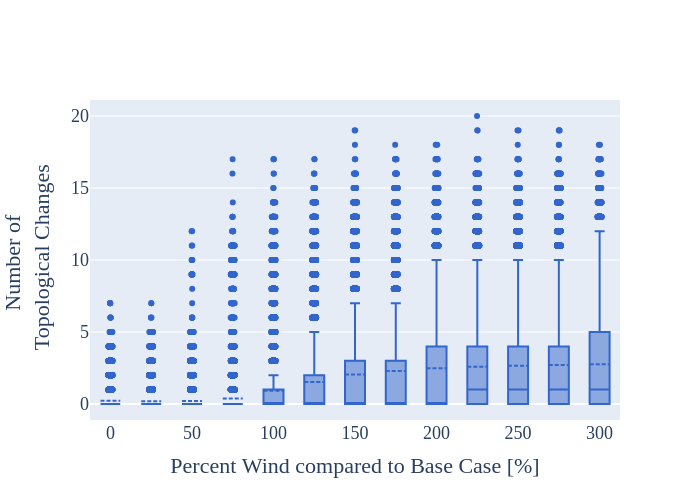}
    \caption{Number of Topology Changes for Each Scenario}
    \label{fig:numTopoChanges}
\end{figure}

We first analyze the effect of increasing renewable energy penetration on the usage of transmission switching. As the wind power generation increases from scenario to scenario, there is an increase in the number of topological changes that are used, shown in Figure \ref{fig:numTopoChanges}. Below the 50\%-Scenario, a maximum of 7 changes occur between any two hours. This goes up to around 20 maximum changes between time steps for cases with more variable generation. Indeed, the scenarios from the lowest wind scenario to the base case level have no changes between time steps more than 75\% of the time. This reduces to around 50\% of the time around the 200\%-Scenario and stays relatively steady for the cases with higher wind generation. It is important to note that while the variability of the topology increases quite quickly between the 50\%-Scenario and the 150\%-Scenario, the increase in topological changes hits an inflection point and begins to level off after a certain threshold. 

\begin{figure}
    \centering
    \includegraphics[width=8.8cm]{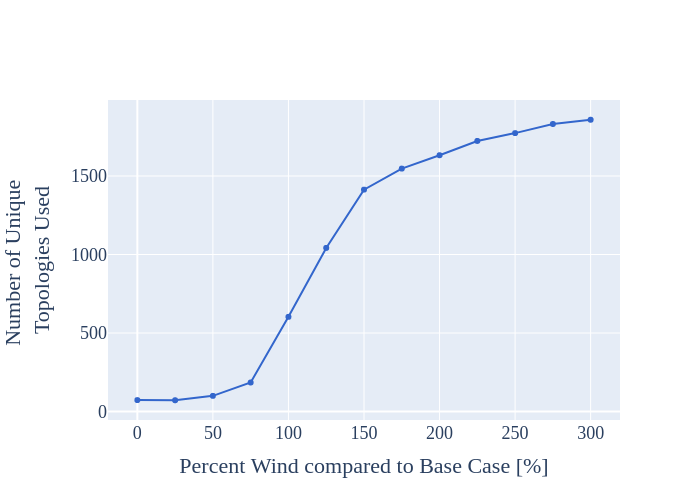}
    \caption{Number of Unique Topologies for Each Scenario}
    \label{fig:numTopo}
\end{figure}

This same threshold is also noticeable in the examination of the number of unique topologies used in each system shown in Figure \ref{fig:numTopo}. This plot shows the number of unique combinations of breaker positions in the system. The number of different topologies increases quickly between the 75\%-Scenario and the 150\%-Scenario, but the number of topologies used increases by less and less after this point. This, along with the results shown in Figure \ref{fig:numTopoChanges} is key as the number of actions that can be used in practice is limited. Figure \ref{fig:numTopoChanges} demonstrates the fact that the ideal number of actions enacted by a system operator between time steps remains fairly steady above a certain wind penetration level. While the number of topologies increases, even in the 300\%-Scenario, the base topology (all lines closed) is used almost half of the hours of the year.

\subsection{Economic Benefits of Topology Control Actions}

\begin{figure} 
    \centering
    \includegraphics[width=8.8cm]{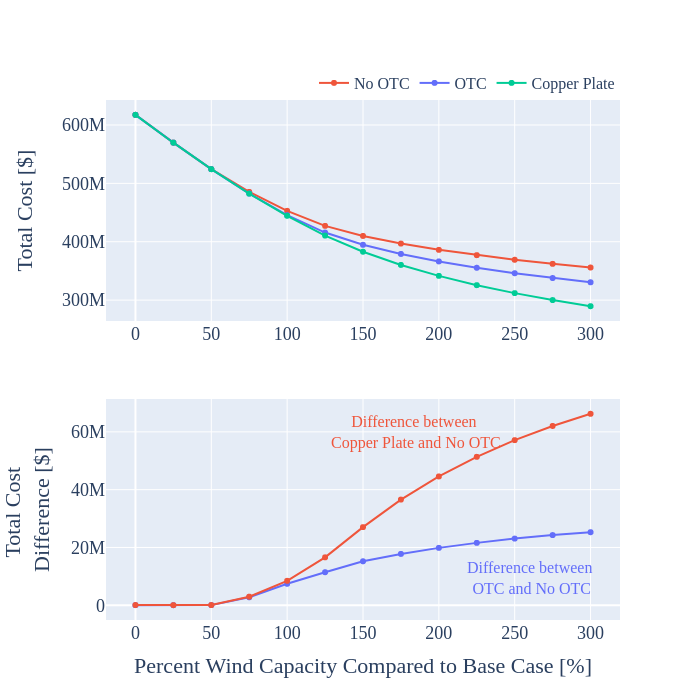}
    \caption{Total Cost Gained due to Topological Control Actions}
    \label{fig:totalCost}
\end{figure}

Figure \ref{fig:totalCost} compares the total cost in the three different situations: copper plate, base network with no grid flexibility and base network with optimal topology control across the various renewable capacity scenarios. The top plot demonstrates the decrease in production cost for each of the three variations as the amount of renewables increases. The difference between the copper plate and the results without topology control are shown in the top line (red) in the bottom plot. The difference in annual production cost between the two variations moves from \$40 thousand to \$66 million as more lower cost energy sources are included in the system. This is over 18\% difference. This value comes solely from the network constraints of the system and gives an upper limit to the possible benefits gained through flexibility. The lower line (blue) in the bottom plot gives the difference between the optimal topology control and the base network without grid flexibility. While the largest rate of change occurs between the 50\% and 150\% scenarios, the gains from the use of grid flexibility continue to increase throughout all scenarios, up to \$20 Million, over 7\% of the total production cost. As the wind production increases by a factor of 3 (between the 100\%- and 300\%-Scenarios), the amount gained due to OTC increases by a factor of 3.4.

\begin{figure}
    \centering
    \includegraphics[width=8.8cm]{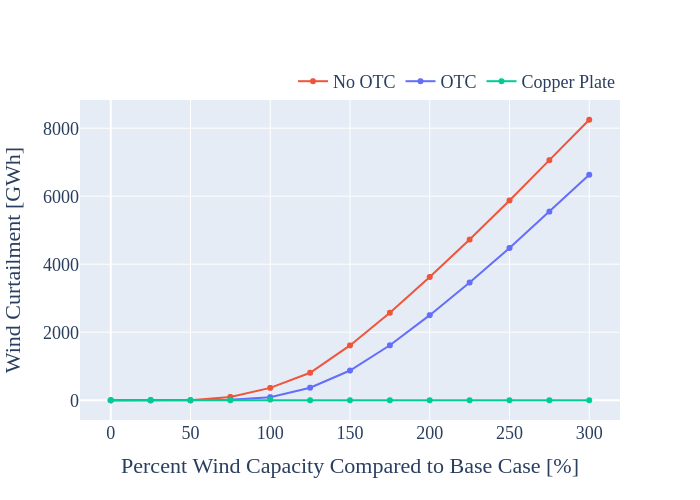}
    \caption{Sum of Wind Curtailment Across All Hours with Thermal Generation}
    \label{fig:curtailment}
\end{figure}

Already the reduction in system cost due to OTC shown in Figure \ref{fig:totalCost} alludes to the fact that a larger amount of renewables are dispatched through the use of grid topology actions. Figure \ref{fig:curtailment} shows the wind curtailment on the hours where there is thermal generation--in other words, it does not include the hours where demand was met with only renewable energy. As a consequence, the copper plate case is shown to have zero wind curtailment for all scenarios, while any curbing of wind generation in the other two cases is due solely to network constraints. The grid flexibility actions allow for a reduction in curtailment of up to 1.6 TWh. The difference in the curves continues to increase as well as more wind is introduced into the system. 

\begin{figure} 
    \centering
    \includegraphics[width=8.8cm]{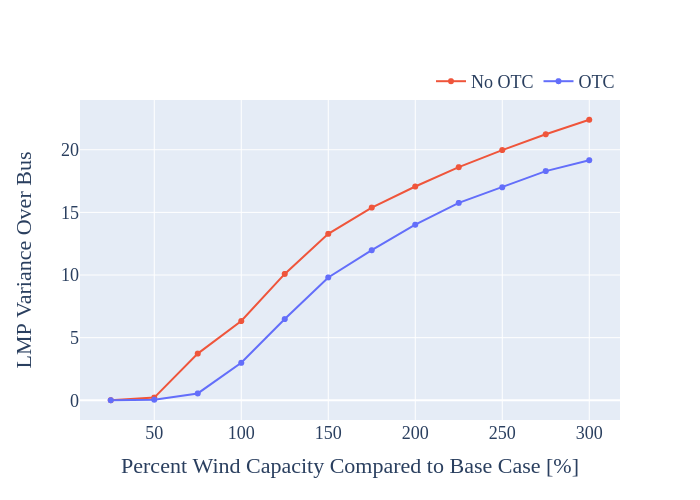}
    \caption{Nodal Price Variance}
    \label{fig:lmp}
\end{figure}

Figure \ref{fig:lmp} shows the average variance of the nodal prices across the buses of the system. This value can be used as an indicator of the congestion in the system. The case with optimal topology control shows a lower geographical variance as the amount of congestion in the system is reduced. 

\section{Conclusion}
This research analyzes how the use of optimal topology controls facilitate the inclusion of renewable energies. As shown through the analysis of a case study, optimal topology control impact on dispatch and cost becomes more important as the energy system grows more variable. While the ideal number of topological actions used increases with the penetration of wind generation, the economic benefit from their use increases along the same axis. In fact, an increase in wind production of a factor of 3 leads to a gain in total system cost by a factor of 3.4 from optimal use of topological controls.

Traditionally, there has been some pushback to market integration of topology optimization especially in nodal markets where switching a line or splitting a bus could render the FTR (Financial Transmission Rights) markets revenue inadequate\cite{hedman_optimal_2011}.  These results show that there will be more and more situations where system operators will be incentivized to use these mechanisms, whether for system security or economic reasons. Their development and integration with existing market designs is thus crucial as variable renewable energies become a larger percent of the dispatch.

\section{Further Research} \label{FurtherResearch}
This research demonstrates the growing value in the optimization of the transmission grid. The next phase of this research is to assess the integration of these topological actions, as well as other grid flexibility measures, into market paradigms. In particular, methods for the inclusion of these levers into the European flow-based market coupling will be assessed.

\bibliographystyle{IEEEtran}

%
\bibliography{bibliography.bib}

\begin{thebibliography}{10}
\providecommand{\url}[1]{#1}
\csname url@samestyle\endcsname
\providecommand{\newblock}{\relax}
\providecommand{\bibinfo}[2]{#2}
\providecommand{\BIBentrySTDinterwordspacing}{\spaceskip=0pt\relax}
\providecommand{\BIBentryALTinterwordstretchfactor}{4}
\providecommand{\BIBentryALTinterwordspacing}{\spaceskip=\fontdimen2\font plus
\BIBentryALTinterwordstretchfactor\fontdimen3\font minus
  \fontdimen4\font\relax}
\providecommand{\BIBforeignlanguage}[2]{{%
\expandafter\ifx\csname l@#1\endcsname\relax
\typeout{** WARNING: IEEEtran.bst: No hyphenation pattern has been}%
\typeout{** loaded for the language `#1'. Using the pattern for}%
\typeout{** the default language instead.}%
\else
\language=\csname l@#1\endcsname
\fi
#2}}
\providecommand{\BIBdecl}{\relax}
\BIBdecl

\bibitem{hedman_review_2011}
\BIBentryALTinterwordspacing
K.~W. Hedman, S.~S. Oren, and R.~P. O'Neill, ``\BIBforeignlanguage{en}{A review
  of transmission switching and network topology optimization},'' in
  \emph{\BIBforeignlanguage{en}{2011 {IEEE} {Power} and {Energy} {Society}
  {General} {Meeting}}}.\hskip 1em plus 0.5em minus 0.4em\relax Detroit, MI,
  USA: IEEE, Jul. 2011, pp. 1--7. [Online]. Available:
  \url{http://ieeexplore.ieee.org/document/6039857/}
\BIBentrySTDinterwordspacing

\bibitem{djelassi_hierarchical_2018}
\BIBentryALTinterwordspacing
H.~Djelassi, S.~Fliscounakis, A.~Mitsos, and P.~Panciatici,
  ``\BIBforeignlanguage{en}{Hierarchical {Programming} for {Worst}-{Case}
  {Analysis} of {Power} {Grids}},'' in \emph{\BIBforeignlanguage{en}{2018
  {Power} {Systems} {Computation} {Conference} ({PSCC})}}.\hskip 1em plus 0.5em
  minus 0.4em\relax Dublin, Ireland: IEEE, Jun. 2018, pp. 1--7. [Online].
  Available: \url{https://ieeexplore.ieee.org/document/8444136/}
\BIBentrySTDinterwordspacing

\bibitem{van_den_bergh_dc_2014}
K.~Van~den Bergh, E.~Delarue, and W.~D'haeseleer, ``Dc power flow in unit
  commitment models,'' May 2014.

\bibitem{mazi_corrective_1986}
A.~A. Mazi, B.~F. Wollenberg, and M.~H. Hesse, ``Corrective {Control} of
  {Power} {System} {Flows} by {Line} and {Bus}-{Bar} {Switching},'' \emph{IEEE
  Transactions on Power Systems}, vol.~1, no.~3, pp. 258--264, Aug. 1986.

\bibitem{van_amerongen_security_1980}
R.~Van~Amerongen and H.~Van~Meeteren, ``Security {Control} by {Real} {Power}
  {Rescheduling}, {Network} {Switching} and {Load} {Shedding}.''\hskip 1em plus
  0.5em minus 0.4em\relax Paris: Cigré, 1980.

\bibitem{lehmann_complexity_2014}
\BIBentryALTinterwordspacing
K.~Lehmann, A.~Grastien, and P.~Van~Hentenryck, ``The {Complexity} of
  {DC}-{Switching} {Problems},'' \emph{arXiv:1411.4369 [cs, math]}, Nov. 2014,
  arXiv: 1411.4369. [Online]. Available: \url{http://arxiv.org/abs/1411.4369}
\BIBentrySTDinterwordspacing

\bibitem{rolim_study_1999}
J.~Rolim and L.~Machado, ``A study of the use of corrective switching in
  transmission systems,'' \emph{IEEE Transactions on Power Systems}, vol.~14,
  no.~1, pp. 336--341, Feb. 1999.

\bibitem{bakirtzis_incorporation_1987}
A.~G. Bakirtzis and A.~P.~S. Meliopoulos, ``Incorporation of {Switching}
  {Operations} in {Power} {System} {Corrective} {Control} {Computations},''
  \emph{IEEE Transactions on Power Systems}, vol.~2, no.~3, pp. 669--675, Aug.
  1987.

\bibitem{schnyder_security_1990}
G.~Schnyder and H.~Glavitsch, ``Security enhancement using an optimal switching
  power flow,'' \emph{IEEE Transactions on Power Systems}, vol.~5, no.~2, pp.
  674--681, May 1990.

\bibitem{bacher_network_1986}
R.~Bacher and H.~Glavitsch, ``Network {Topology} {Optimization} with {Security}
  {Constraints},'' \emph{IEEE Transactions on Power Systems}, vol.~1, no.~4,
  pp. 103--111, Nov. 1986.

\bibitem{koglin_corrective_1985}
\BIBentryALTinterwordspacing
H.-J. Koglin and M.~de~Medeiros, ``\BIBforeignlanguage{en}{Corrective
  {Switching} {Approaching} {On}-{Line} {Application}},''
  \emph{\BIBforeignlanguage{en}{IFAC Proceedings Volumes}}, vol.~18, no.~7, pp.
  203--207, Jul. 1985. [Online]. Available:
  \url{https://linkinghub.elsevier.com/retrieve/pii/S1474667017604361}
\BIBentrySTDinterwordspacing

\bibitem{bertram_integrated_1990}
T.~Bertram, K.~Demaree, and L.~Dangelmaier, ``An integrated package for
  real-time security enhancement,'' \emph{IEEE Transactions on Power Systems},
  vol.~5, no.~2, pp. 592--600, May 1990.

\bibitem{gorenstin_b_framework_1987}
{Gorenstin, B.}, {Terry, L. A.}, {Pereira, M. V. F.}, and {Pinto, L. M. V. G.},
  ``A framework for integration of network topology optimization and generation
  rescheduling in power system security applications,'' in \emph{System
  security and optimization {I}}.\hskip 1em plus 0.5em minus 0.4em\relax
  Cascais, Portugal: Butterworths, Sep. 1987, pp. 124--130.

\bibitem{freitas_e_silva_switching_1993}
J.~a.~O. Freitas~e Silva and L.~J.~B. Machado, ``Switching {Lines} {Selection}
  to {Integrate} the {Network} {Topology} {Optimization} with the {Usual}
  {Overload} {Control} {Actions} of {Electric} {Power} {Systems},'' Avignon,
  France, 1993.

\bibitem{makram_selection_1989}
E.~B. Makram, K.~P. Thornton, and H.~E. Brown, ``Selection of {Lines} to {Be}
  {Switched} to {Eliminate} {Overloaded} {Lines} {Using} a {Z}-{Matrix}
  {Method},'' \emph{IEEE Power Engineering Review}, vol.~9, no.~5, pp. 63--63,
  May 1989.

\bibitem{rolim_secte_1995}
\BIBentryALTinterwordspacing
J.~G. Rolim, M.~R. Irving, and L.~J.~B. Machado, ``{SECTE} - {An} {Expert}
  {System} {For} {Voltage} {Control}, {Including} {Topological} {Changes},''
  \emph{IFAC Proceedings Volumes}, vol.~28, no.~26, pp. 177 -- 182, 1995.
  [Online]. Available:
  \url{http://www.sciencedirect.com/science/article/pii/S1474667017447537}
\BIBentrySTDinterwordspacing

\bibitem{dodu_search_1981}
J.~Dodu, A.~Merlin, and J.~David, ``On the search of optimal switching
  configurations in power transmission systems studies,'' in \emph{Power
  {System} {Computation} {Conference} ({PSCC})}, 1981, pp. 282--292.

\bibitem{glavitsch_h_combined_1984}
{Glavitsch, H.}, {Kronig, H.}, and {Bacher, R.}, ``Combined use of linear
  programming and load flow techniques in determining optimal switching
  sequences,'' in \emph{Network {Adequacy}}.\hskip 1em plus 0.5em minus
  0.4em\relax Helsinki, Finland: Butterworths, Aug. 1984, pp. 627--636.

\bibitem{oneill_dispatchable_2005}
\BIBentryALTinterwordspacing
R.~O'Neill, R.~Baldick, U.~Helman, M.~Rothkopf, and W.~Stewart~Jr.,
  ``\BIBforeignlanguage{en}{Dispatchable {Transmission} in {RTO} {Markets}},''
  \emph{\BIBforeignlanguage{en}{IEEE Transactions on Power Systems}}, vol.~20,
  no.~1, pp. 171--179, Feb. 2005. [Online]. Available:
  \url{http://ieeexplore.ieee.org/document/1388507/}
\BIBentrySTDinterwordspacing

\bibitem{fisher_optimal_2008}
\BIBentryALTinterwordspacing
E.~Fisher, R.~O'Neill, and M.~Ferris, ``\BIBforeignlanguage{en}{Optimal
  {Transmission} {Switching}},'' \emph{\BIBforeignlanguage{en}{IEEE
  Transactions on Power Systems}}, vol.~23, no.~3, pp. 1346--1355, Aug. 2008,
  38. [Online]. Available: \url{http://ieeexplore.ieee.org/document/4492805/}
\BIBentrySTDinterwordspacing

\bibitem{hedman_co-optimization_2009}
K.~W. Hedman and S.~S. Oren, ``\BIBforeignlanguage{en}{Co-optimization of
  {Generation} {Unit} {Commitment} and {Transmission} {Switching} with {N}-1
  {Reliability}},'' p.~12, Sep. 2009.

\bibitem{hedman_optimal_2009}
K.~W. Hedman, R.~P. O'Neill, E.~B. Fisher, and S.~S. Oren, ``Optimal
  {Transmission} {Switching} {With} {Contingency} {Analysis},'' \emph{IEEE
  Transactions on Power Systems}, vol.~24, no.~3, pp. 1577--1586, Aug. 2009.

\bibitem{goldis_shift_2017}
E.~A. Goldis, P.~A. Ruiz, M.~C. Caramanis, X.~Li, C.~R. Philbrick, and A.~M.
  Rudkevich, ``Shift {Factor}-{Based} {SCOPF} {Topology} {Control} {MIP}
  {Formulations} {With} {Substation} {Configurations},'' \emph{IEEE
  Transactions on Power Systems}, vol.~32, no.~2, pp. 1179--1190, Mar. 2017.

\bibitem{han_impacts_2016}
\BIBentryALTinterwordspacing
J.~Han and A.~Papavasiliou, ``\BIBforeignlanguage{en}{The {Impacts} of
  {Transmission} {Topology} {Control} on the {European} {Electricity}
  {Network}},'' \emph{\BIBforeignlanguage{en}{IEEE Transactions on Power
  Systems}}, vol.~31, no.~1, pp. 496--507, Jan. 2016. [Online]. Available:
  \url{http://ieeexplore.ieee.org/document/7063277/}
\BIBentrySTDinterwordspacing

\bibitem{hedman_optimal_2008}
K.~W. Hedman, R.~P. O'Neill, E.~B. Fisher, and S.~S. Oren, ``Optimal
  {Transmission} {Switching}—{Sensitivity} {Analysis} and {Extensions},''
  \emph{IEEE Transactions on Power Systems}, vol.~23, no.~3, pp. 1469--1479,
  Aug. 2008.

\bibitem{hedman_optimal_2011}
\BIBentryALTinterwordspacing
K.~W. Hedman, S.~S. Oren, and R.~P. O’Neill,
  ``\BIBforeignlanguage{en}{Optimal transmission switching: economic efficiency
  and market implications},'' \emph{\BIBforeignlanguage{en}{Journal of
  Regulatory Economics}}, vol.~40, no.~2, pp. 111--140, Oct. 2011. [Online].
  Available: \url{http://link.springer.com/10.1007/s11149-011-9158-z}
\BIBentrySTDinterwordspacing

\bibitem{mekonnen_influence_2012}
\BIBentryALTinterwordspacing
M.~T. Mekonnen and R.~Belmans, ``\BIBforeignlanguage{en}{The influence of phase
  shifting transformers on the results of flow-based market coupling},'' in
  \emph{\BIBforeignlanguage{en}{2012 9th {International} {Conference} on the
  {European} {Energy} {Market}}}.\hskip 1em plus 0.5em minus 0.4em\relax
  Florence, Italy: IEEE, May 2012, pp. 1--7. [Online]. Available:
  \url{http://ieeexplore.ieee.org/document/6254746/}
\BIBentrySTDinterwordspacing

\bibitem{mekonnen_power_2013}
\BIBentryALTinterwordspacing
M.~T. Mekonnen, C.~De~Jonghe, B.~Rawn, D.~Van~Hertem, and R.~Belmans,
  ``\BIBforeignlanguage{en}{Power flow control and its effect on flow-based
  transmission cost allocation},'' in \emph{\BIBforeignlanguage{en}{2013 10th
  {International} {Conference} on the {European} {Energy} {Market}
  ({EEM})}}.\hskip 1em plus 0.5em minus 0.4em\relax Stockholm, Sweden: IEEE,
  May 2013, pp. 1--8. [Online]. Available:
  \url{http://ieeexplore.ieee.org/document/6607388/}
\BIBentrySTDinterwordspacing

\bibitem{oneill_efficient_2005}
\BIBentryALTinterwordspacing
R.~P. O'Neill, P.~M. Sotkiewicz, B.~F. Hobbs, M.~H. Rothkopf, and W.~R.
  Stewart, ``\BIBforeignlanguage{en}{Efficient market-clearing prices in
  markets with nonconvexities},'' \emph{\BIBforeignlanguage{en}{European
  Journal of Operational Research}}, vol. 164, no.~1, pp. 269--285, Jul. 2005.
  [Online]. Available:
  \url{https://www.sciencedirect.com/science/article/abs/pii/S0377221703009196}
\BIBentrySTDinterwordspacing

\bibitem{barrows_ieee_2020}
\BIBentryALTinterwordspacing
C.~Barrows, E.~Preston, A.~Staid, G.~Stephen, J.-P. Watson, A.~Bloom, A.~Ehlen,
  J.~Ikaheimo, J.~Jorgenson, D.~Krishnamurthy, J.~Lau, B.~McBennett, and
  M.~O'Connell, ``\BIBforeignlanguage{en}{The {IEEE} {Reliability} {Test}
  {System}: {A} {Proposed} 2019 {Update}},'' \emph{\BIBforeignlanguage{en}{IEEE
  Transactions on Power Systems}}, vol.~35, no.~1, pp. 119--127, Jan. 2020.
  [Online]. Available: \url{https://ieeexplore.ieee.org/document/8753693/}
\BIBentrySTDinterwordspacing

\bibitem{grigg_ieee_1999}
C.~Grigg, P.~Wong, P.~Albrecht, R.~Allan, M.~Bhavaraju, R.~Billinton, Q.~Chen,
  C.~Fong, S.~Haddad, S.~Kuruganty, W.~Li, R.~Mukerji, D.~Patton, N.~Rau,
  D.~Reppen, A.~Schneider, M.~Shahidehpour, and C.~Singh, ``The {IEEE}
  {Reliability} {Test} {System}-1996. {A} report prepared by the {Reliability}
  {Test} {System} {Task} {Force} of the {Application} of {Probability}
  {Methods} {Subcommittee},'' \emph{IEEE Transactions on Power Systems},
  vol.~14, no.~3, pp. 1010--1020, Aug. 1999.

\end{thebibliography}
 
\end{document}